\documentclass[
letterpaper, 
10pt,
final,
twocolumn,
superscriptaddress,
floats,
floatfix,
longbibliography,
aps,
pra
]{revtex4-1}
\usepackage{mathtools} 
\usepackage{amsmath}
\usepackage{amssymb}
\usepackage{makeidx}
\usepackage{acronym}
\usepackage{natbib}
\usepackage[version=3]{mhchem}
\usepackage{color}
\usepackage{booktabs}
\usepackage{graphicx}
\usepackage{caption}
\usepackage{subcaption}
\usepackage{rotating}
\usepackage[linewidth=1pt]{mdframed}

\newcommand{\adhoc}{\textit{ad hoc}}

\newcommand{\via}{\textit{via}}
\newcommand{\vs}{\textit{vs}}

\newcommand{\eg}{{e.g.}, }
\newcommand{\ie}{{i.e.}, }

\begin{document}
\author{Gaurav Vishwakarma}
\email{gvishwak@buffalo.edu}
\affiliation{Department of Chemical and Biological Engineering, University at Buffalo, The State University of New York, Buffalo, NY 14260, United States}
\author{Aditya Sonpal}
\affiliation{Department of Chemical and Biological Engineering, University at Buffalo, The State University of New York, Buffalo, NY 14260, United States}
\author{Johannes Hachmann}
\email{hachmann@buffalo.edu}
\affiliation{Department of Chemical and Biological Engineering, University at Buffalo, The State University of New York, Buffalo, NY 14260, United States}
\affiliation{Computational and Data-Enabled Science and Engineering Graduate Program, University at Buffalo, The State University of New York, Buffalo, NY 14260, United States}
\affiliation{New York State Center of Excellence in Materials Informatics, Buffalo, NY 14203, United States}

\title{Metrics for Benchmarking and Uncertainty Quantification: Quality, Applicability, and a Path to Best Practices for Machine Learning in Chemistry}

\begin{abstract} 
This review aims to draw attention to two issues of concern when we set out to make machine learning work in the chemical and materials domain, \ie\ statistical loss function metrics for the validation and benchmarking of data-derived models, and the uncertainty quantification of predictions made by them. 
They are often overlooked or underappreciated topics as chemists typically only have limited training in statistics.
Aside from helping to assess the quality, reliability, and applicability of a given model, these metrics are also key to comparing the performance of different models and thus for developing guidelines and best practices for the successful application of machine learning in chemistry. 
\end{abstract}


\maketitle


\section{Assessing Machine Learning Models}
\label{sec:intro}
The rapid advancement and transformation of machine learning (ML) technology has led to a boom in its utilization, including in science and engineering. 
Chemical research is no longer an exception in this development \cite{nsfreport}, and numerous areas have been identified, in which ML is now employed to great effect (see, \eg\ Refs.\ \cite{haghighatlari2019advances,Afzal2019d,Afzal2019c,Afzal2018a,Haghighatlari2019d,Haghighatlari2019c}). 
While ML applications have resulted in a number of exciting and valuable studies that have advanced chemical domain knowledge, it is worth noting that there is still a considerable lack of quality control, guidance, uniformity, and established protocols for the successful conduct of such studies. 
Unlike for other application domains of ML or for other techniques used in chemistry, there are no decades of experience to build on. 
Guidelines established in other contexts do not necessarily translate to chemical problem settings.

The choices that define chemical ML models, \eg\ with respect to featurization (balancing expressiveness and cost), training data sampling (accounting for data volume limitations, biases, imbalances), ML hyperparameter and model selection (balancing complexity and effectiveness), etc., have a dramatic impact on the resulting models' predictive performance and range of applicability. 
So far, the community has mostly relied on \adhoc\ choices that are unlikely to yield the best possible outcomes.
The ability to quantify the quality, reliability, and applicability of ML models \via\ metrics is thus an obvious topic of interest.
ML approaches that optimize the model design choices do so by minimizing an error metric (\eg\ \via\ a fitness function in an evolutionary algorithm \cite{vishwakarma2019towards}).
The comparison of different models on the basis of these metrics can also yield design recommendations, illuminate their implications, and thus result in best practices for different problem scenarios within the chemistry domain. 
Ultimately, they may serve as the foundation for meta-ML facilities and expert recommender systems as part of ML software tools (\eg\ Ref.\ \cite{Hachmann2018,haghighatlari2019chemml,Hanwell2020,gunawardana2009survey}).

\section{Pieces of the Metrics Puzzle}
\label{sec:overview}
For ML regression and classification models, there are numerous statistical metrics (also known as loss function metrics) that can be used to characterize their  performance.
The notion of 'no-free-lunch' \cite{wolpert2005coevolutionary} in computational complexity and optimization theorizes that the performance of any two methods or algorithms is equivalent when averaged across all possible problems. 
This theorem applies to various aspects of both model selection and validation in ML as well \cite{makridakis1993accuracy}. 
Loss function metrics are generally based on the comparison of model predictions $y_{i, pred}$ and an assumed ground truth $y_{i, true}$ for a number of instances $i$, which leads to prediction errors $e_{i}$ (Eqn.\ \ref{ei}) and relative prediction errors $r_{i}$ (Eqn.\ \ref{ri}), respectively.

Different metrics illuminate different performance aspects of a model. 
A clear understanding of the specific information a given metric conveys is a prerequisite to fully harnessing it. 
Blind reliance on a random (\eg\ default or commonly reported) metric is a missed opportunity at best and leads to poor outcomes at worst.
While particular metrics may be of greater or lesser importance for different application problems, it is generally worth considering a compilation of metrics. 
Individual metrics only yield limited insights and no single metric by itself can fully capture the performance of an ML model. 
But taken together, different metrics complement each other and -- like pieces of a puzzle -- paint a comprehensive picture of a model's quality.

The same metrics with respect to the same ground truth need to be compared between different models or studies, otherwise the comparison is meaningless.
As an alternative to comparing the error metrics of two models (with respect to an independent ground truth), we can also choose the ground truth to be the predictions of one of the models. In that case, the error metrics directly reflect the differences between the two models.

It is important to stress that while error metrics can be applied to the predictions within the training, validation, and test data set (including as part of $k$-fold cross-validation, in which these sets get reshuffled), only the results for the unseen test set data is considered in the evaluation of the predictive performance of a model. The comparison of training and test set error metrics is instructive as significant differences indicate a poorly trained (\eg\ overfitted) model. Similarly, the errors of the different instances of a $k$-fold cross-validation should be consistent.

In the following sections, we will provide a concise overview of a selection of particularly useful metrics, highlight their advantages and disadvantages, and discuss how a suite of these metrics can afford multifaceted insights into the behavior of a model. It is worth mentioning that much of this discussion is transferable to predictions of non-ML (\eg\ physics-based rather than data-derived) models \cite{Afzal2018b}.
We also stress that all prediction errors have to be judged in the context of the intrinsic errors or uncertainties of the assumed ground truth.

\section{Metrics for Model Validation and Benchmarking}
\label{sec:validation}
\subsection{Regression Tasks}
For regression tasks, the \textit{mean absolute error} (MAE) and \textit{root mean square error} (RMSE) are two of the most commonly reported error metrics (Eqns.\ \ref{mae}, \ref{rmse}), and a number of studies have been published debating the supremacy of one over the other \cite{willmott2005advantages, chai2014root, wang2018analysis, brassington2017mean, willmott2006use, willmott1982some, armstrong1992error, Pernot_2020, pernot_sip}. 
(Note that \textit{mean absolute deviation} (MAD) and \textit{root mean square deviation} (RMSD) are sometimes used synonymously with MAE and RMSE, respectively. 
However, since these abbreviations are also used for other statistical metrics such as median absolute deviation or with other definitions, 
we do not recommend their use to avoid confusion or erroneous conclusions.)
The MAE (also called \textit{mean unsigned error} (MUE)) provides straightforward information about the average magnitude of errors to be expected from a model. 
However, as all errors are weighted equally, differences in the magnitudes of errors get averaged out, \ie\ the MAE alone does not offer insights into the uniformity or variability of prediction errors (and thus the reliability of particular predictions).
Metrics that rely on squared errors, such as the RMSE or the less frequently reported \textit{mean square error} (MSE), magnify larger errors and are thus more sensitive to outliers (which are signaled by large RMSE values).
Considered together, MAE and RMSE can yield information on the homogeneity or heterogeneity of errors: if MAE and RMSE values are similar, this indicates prediction errors of relatively consistent magnitude; if the RMSE is significantly larger than the MAE, this indicates large fluctuations in the error magnitudes \cite{syntetos2005accuracy}.

MAE and RMSE provide absolute errors that are decoupled from the prediction values. However, the same absolute error has very different implications for smaller or larger prediction values.   
The \textit{mean absolute percentage error} (MAPE) and \textit{root mean square percentage error} (RMSPE) given by Eqns.\ \ref{mape} and \ref{rmspe}, respectively, provide error metrics that are relative to the prediction values, and thus complement the absolute MAE and RMSE values. 
The comparison of MAPE and RMSPE allow us to gauge the uniformity of prediction errors across the range of prediction values (rather than their absolute uniformity; 
note that absolute and relative uniformity will generally not be achievable at the same time, unless the range of prediction values is very narrow). 
Use-cases are limited to non-zero prediction values \cite{swanson2011mape, ren2009applicability, kolassa2011percentage, goodwin1999asymmetry}.

The unsigned errors discussed so far only consider error magnitudes, but not their directional distribution around the prediction. 
The \textit{mean error} (ME) and \textit{mean percentage error} (MPE) given by Eqns.\ \ref{me} and \ref{mpe}, respectively, allow us to identify systematic biases in the directionality of errors. 
Unbiased absolute and relative errors have ME and MPE values of 0.0. Positive ME and MPE values indicate systematic overpredictions and negative ones systematic underpredictions. Their magnitude corresponds to the degree of directional bias.

All metrics considered so far provide average errors. They can be complemented by the \textit{maximum absolute error} (MaxAE) and \textit{maximum absolute percentage error} (MaxAPE) given by Eqns.\ \ref{maxae} and \ref{maxape}, respectively, as well as the difference of most extreme errors $\Delta$MaxE (Eqn.\ \ref{deltamaxe}), \ie\ the \textit{spread} between largest positive and negative errors. 
These three metrics provide absolute and relative worst cases in the observed prediction errors. 
Comparing the maximum error metrics with their corresponding means indicates the degree of deviation between them.

We can further characterize the absolute and/or relative prediction error distributions. Ideally, these should be normal distributions centered around 0.0 with narrow \textit{standard deviations} $\sigma$ (Eqn.\ \ref{sigma}), \ie\ the square root of the \textit{variance} $\sigma^2$. The center of the error and percentage error distributions are ME and MPE, respectively. A negligible directional bias means that a method is \textit{accurate}, while small $\sigma$ means that a method is \textit{precise}.

\begin{figure*}[t]
    \centering
    \includegraphics[width=0.95\textwidth]{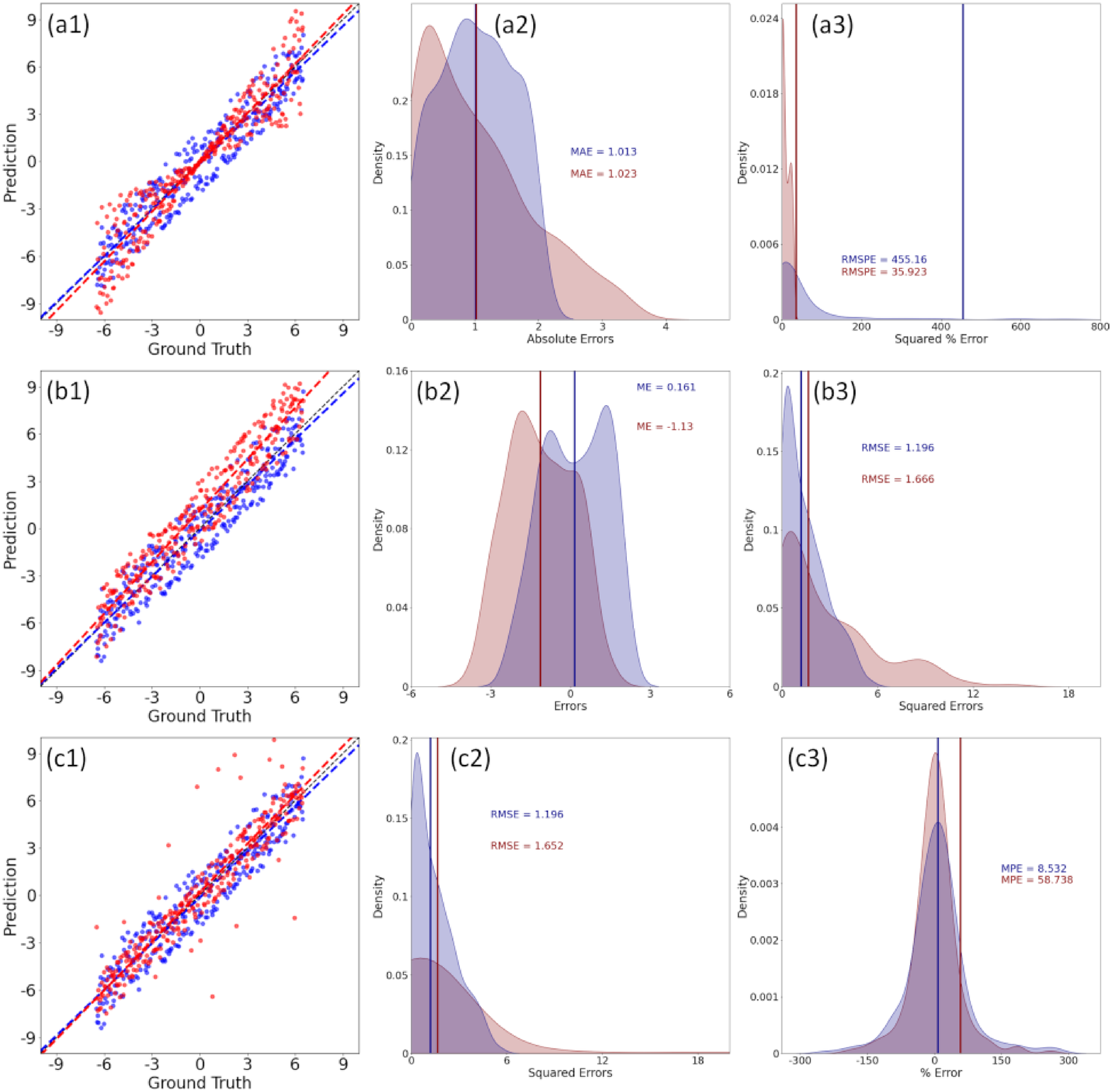}
    \caption{This figure compares pairs of data sets with different error characteristics and how these are reflected in different error metrics. The data sets each contain 300 data points that are synthetically generated using $y_i = rn(k|l) \cdot x_i + rn(m|n)$, with linear and constant random noise $rn$ within the ranges $(k|l)$ and $(m|n)$, respectively. We use these ranges to simulate different error scenarios. Column 1 shows plots that compare the `predictions' $y$, with the ground truth parity line $y = x$ in black. Linear regression lines for each data set are shown in their respective color. Columns 2 and 3 contain error density plots for selected metrics. (a1) shows data sets with only linear (red) or only constant errors (blue), chosen such that their MAE values are near identical (a2). While the red data has a slightly larger RMSE, the blue data has dramatically larger MAPE and RMSPE (a3). 
For the red data, MAE and RMSE are similar and MAPE and RMSPE dissimilar, while the situation for the blue data is reversed. 
(b1) shows data sets with a mix of linear and constant error, one where these are directionally biased (red), the other unbiased (blue). The bias in red is reflected in significantly non-zero ME (b2) and an incorrect regression slope.
As the red data follows the wrong trend, all error metrics are elevated, \eg\ RMSE (b3). 
In (c1), the blue data is the same as in (b1). In the red data, 
10 \% of the data points are replaced with outliers. The error of the remaining data is reduced such that the overall MAE of both sets match. The outliers result in increased RMSE (c2), comparable to that seen for systematic errors (b3). 
They also by chance led to large non-zero MPE values (c3) that could easily be mistaken for a sign of systematic bias. However, the slope and ME values readily disprove this notion.}
\label{f:dummy_data}  
\end{figure*}

We can also quantify the extent of correlation between the prediction results and ground truth by performing a linear regression. 
The \textit{coefficient of determination} $R^2$ (Eqn.\ \ref{r2}), with $R$ the \textit{correlation coefficient}, of the fit is a widely reported metric.
Maximizing $R^2$ towards the upper limit of 1.0 is equivalent to minimizing the MSE.
The \textit{slope} and \textit{offset} values of the linear regression (\ie\ deviations from 1.0 for the former and 0.0 for the latter) yield additional insights about systematic error behavior that can complement our findings from the ME, MPE, and $\sigma$ metrics. 
Instead of the $R^2$ value, some studies report the \textit{adjusted coefficient of determination} $R^2_{adj}$ (Eqn.\ \ref{adjr2}), which incorporates a measure of model complexity, thus giving information about the quality/complexity ratio. 
While the $R^2$ increases monotonously with the number of features or variables added to a model, 
the $R^2_{adj}$ increases only when useful features are added, and decreases otherwise.
We could in principle also perform non-linear regressions to further explore the nature of systematic biases, but this is in practice rarely done, as the need for such metrics suggests more fundamental flaws in our ML model. 
Instead, we could employ $\Delta$-ML or transfer learning techniques to directly correct for the discrepancies between model predictions and ground truth and thus augment and improve the original ML model. 

Fig.\ \ref{f:dummy_data} shows the characteristics and utility of several of the regression metrics discussed in this section for different types of errors.

In summary, a good ML model should make predictions with small MAE, RMSE, MAPE, and RMSPE values; small differences between either MAE and RMSE (\ie\ homogeneous absolute errors) or MAPE and RMSPE (\ie\ homogeneous relative errors); ME and MPE values close to 0.0; small $\sigma$; small MaxAE and MaxAPE values with only modest differences to MAE and MAPE, respectively; small $\Delta$MaxE value; $R^2$ and slope close to 1.0 and offset close to 0.0.



\subsection{Classification Tasks}
\begin{figure}[t] 
	\centering
	\includegraphics[width=0.47\textwidth]{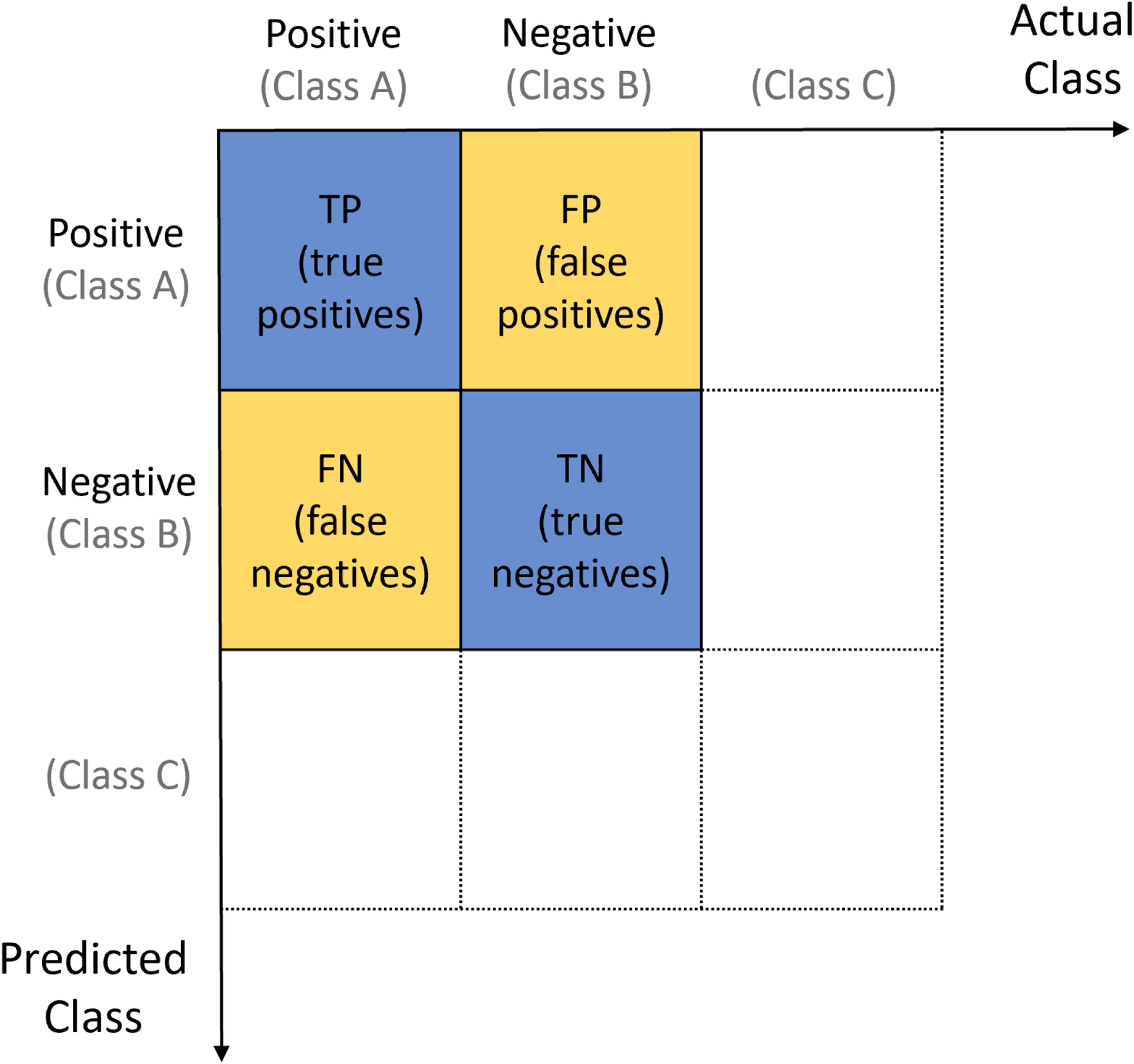}
	\caption{Confusion matrix for binary classification problems. The matrix can be extended for multi-class classifications where positive and negative classes are replaced with classes A, B, C, \textit{etc}.}
	\label{fig:conf}
\end{figure}

A simple way of visualizing and reporting the quality of results for classification tasks is \via\ a confusion matrix (Fig.\ \ref{fig:conf}) \cite{STEHMAN199777}, which can be used for both binary and multi-class classifications. 
A confusion matrix is a square matrix (of size equal to the number of classes) that represents a model's performance by tabulating class-specific information about the number of correct and incorrect predictions. 
For a binary classification task, a confusion matrix shows the \textit{total number of true positive} (TP), \textit{true negative} (TN), \textit{false positive} (FP), and \textit{false negative} (FN) predictions. 
These values can be used to calculate other evaluation metrics, including accuracy, precision, and recall.

The simplest of these derived metrics is the \textit{accuracy}, which is defined as the fraction of correctly labelled predictions among the total number of cases examined (Eqn.\ \ref{acc}). 
While this metric is easy to interpret and suitable for binary and multi-class classification alike, it falls short when dealing with skewed or imbalanced data \cite{lavravc1999rule, gu2009evaluation, hossin2011novel}. 
For cases where the data set is not necessarily balanced, metrics such as precision and recall are preferred \cite{furnkranz2003analysis}. 
In binary classification problems, \textit{precision} denotes the fraction of positive class labels that are predicted correctly by the model (Eqn.\ \ref{prec}). 
\textit{Recall} denotes the overall fraction of the positive class labels that are correctly predicted (Eqn.\ \ref{rec}). 
It is preferred when false negatives are highly undesirable (\eg\ if a toxic chemical is falsely predicted to be non-toxic, then it will have far greater ramifications than if a non-toxic chemical is classified as toxic). 
Thus, in situations where the negative class represents an overwhelming fraction of the training data, precision and recall are more useful than accuracy since it is imperative that all data points belonging to the positive class are predicted correctly. 
Accuracy, precision, and recall values close to the upper limit of 1.0 are indicative of a well-performing model.

In most cases, there is a trade-off between precision and recall. 
The \textit{F1 score}, which is the harmonic mean of precision and recall (Eqn.\ \ref{f1}), is a useful metric when it is desirable to have a balance between precision and recall \cite{Powers2008,Baeza-Yates1999}. 
The F1 score gives equal weight to precision and recall, however, when domain knowledge or other considerations indicate that more weight should be assigned to one or the other, we can use a \textit{weighted F1 score} (F1\textsubscript{$\beta$}), which introduces a weight parameter $\beta$ to adjust the precision-recall (PR) trade-off (Eqn.\ \ref{wf1}).

In certain classification problems, the output of a classifier for a given input is a probability distribution over a set of class labels rather than just the most likely class label. 
Metrics used to evaluate predicted probabilities are different from those used to evaluate class labels. 
For predicted probabilities resulting from binary classification, \textit{log loss} $\mathcal{L}$ (also called \textit{binary cross-entropy}) (Eqn.\ \ref{logloss}) is considered a good metric. 
Although it primarily serves as a fitness function for classifiers, it can also be used as an evaluation metric. 
While it successfully accounts for the uncertainty of a model's prediction, it needs to be modified with class weights in case of imbalanced data. 
An extension of this metric for multi-class classifications is the \textit{categorical cross-entropy} \cite{a8943952, gordon2020uses}.

While predicted probabilities give a more nuanced view of a classifier's performance, distinct class labels are preferred for most practical purposes. 
The latter are derived from the former \via\ a threshold. 
Two diagnostic metrics (along with domain knowledge) are commonly used to determine the best threshold value, which in turn determines the balance of the classes in the data set. 
These metrics are the \textit{receiver operating characteristic} (ROC) curve \cite{Fawcett2006roc, ferri2002learning} and the PR curve \cite{takaya2015}. 
The ROC curve is a plot of the \textit{true positive rate} (TPR) (Eqn.\ \ref{rec}) \vs\ the \textit{false positive rate} (FPR) (Eqn.\ \ref{fpr}) at each threshold value. 
(Note that the TPR is the same as the recall.)
The optimum threshold value is one that has a high TPR and a low FPR. 
Given the ROC curve, we can also compute the \textit{area under the ROC curve} (AUC) \cite{Hand2001auc,Huang2005,Rakotomamonjy2004,Flach2003,McClish1989}, which is an important metric used for model selection in classification problems. 
The closer the AUC value is to the upper limit of 1.0, the better a model performs. 
We utilize these metrics by plotting the ROC curve with different thresholds and then comparing the AUC for the optimal threshold values for different models.

One shortcoming of ROC curves is that they do not work well for imbalanced or skewed data \cite{drummond2006cost}. 
For such data sets, PR curves have greater utility \cite{davis2006relationship}. 
A PR curve is a plot of the model's precision \vs\ recall at different threshold values. 
The threshold for which the model has both a high precision as well as a high recall is selected as the optimum value. 
The F1 score at each threshold can also be determined, along with the \textit{area under the PR curve} (which is ideally close to 1.0 for a good classifier) and is used for model selection.

In summary, the choice of metrics to assess the quality of an ML classification model depends on the nature of the given data (\ie\ balanced or imbalanced), application of the model (which determines the weight to be assigned to positive or negative class labels), and the nature of the classifier itself (\ie\ whether it predicts probabilities or individual class labels). 
As discussed before, it is prudent to compute a set of metrics to obtain a well-developed understanding of a model's performance.


\section{Metrics for Uncertainty Quantification}
\label{sec:app_domain} 
Aside from creating and benchmarking an ML model, an equally important task is to ascertain its applicability to a target domain of interest.
For chemistry and drug-related applications, where generally the molecules are numerically represented using fingerprints \cite{Morgan1965, ecfp2010, Carhart1985hap, Ramaswamy1987htt, maccs2002, Taletesrl2011, LandrumRdkit, OBoyle2011}, it is common practice to use similarity metrics such as the \textit{Tanimoto index} T \cite{bajusz2015tanimoto} (also called \textit{Jaccard coefficient}) (Eqn.\ \ref{tanimoto}) to gauge the similarity of target molecules to those in the training set. Similarity in the training and target domains indicates that the predictive performance of the ML model should hold for the target domain.

Formal uncertainty quantification is relatively straightforward if (i) the distribution of the data is known, (ii) the ML model is linear, or (iii) if the model inherently provides an uncertainty for each prediction (such as in Bayesian learning approaches, Gaussian processes, or random forests \cite{meinshausen2006quantile}). 
If these scenarios do not apply, then we can use a number of non-parametric, model-agnostic methods to quantify the reliability of predictions made by ML models for a target or `query' point.
The perhaps best-known method that has successfully been employed in both regression and classification problems is the \textit{ensemble variance} (also known as the \textit{sensitivity analysis}) method. 
In this method, we create an ensemble of ML models by repeatedly sampling (with replacement) subsets of the training data (also known as bootstrap aggregating or bagging \cite{mentch2016quantifying}). The variance in their predictions for a query point is used to determine, whether or not the query point lies within the applicability domain \cite{musil2019fast, peterson2017addressing, bosnic2008comparison, Toplak2014}. 
The smaller the variance in the predictions, the more likely it is that the query point falls into the applicability domain, whereas larger variances are more likely an indication of the query point being an outlier.
Unfortunately, this method has a high computational overhead, in particular with complex models and/or large data, which limits its practical utility.

Another class of methods is based on the \textit{range of descriptor values} (or those of other representations).
For instance, we can examine every descriptor value in the query point with the corresponding range across all points in the training data to assess the applicability of the model to the query point \cite{jaworska2005qsar}. 
In \textit{geometric methods}, we construct convex hulls around the training data to define the extent of the descriptor values. 
These methods have also been extended to data obtained after a transformation of the initial set of descriptors, such as a representation obtained from a principal component analysis (PCA). 
However, insights about the density distribution of descriptor values cannot be inferred from range-based methods.

Finally, we can also employ techniques that are \textit{distance-based}, \ie\ they rely on the distance of the query point from the distribution of the training set, assuming that ML predictions are trustworthy in regions of dense data. 
Distance-based metrics tend to be easy and inexpensive to compute. 
A model's applicability domain is determined \via\ a predefined threshold for the distance of a query point from a point within the distribution. 
This can either be the distance to the mean of the distribution, average (or weighted-average \cite{Liu2019, liu2018molecular}) distance to $k$-nearest neighbors (neighbors with similar descriptor values) in the training set or the maximum or average distance to all of the points in the distribution. 
The \textit{Euclidean} and \textit{Mahalanobis distances} are the most common distance metrics employed to quantify the distance to a distribution of data points. 
The Mahalanobis distance indicates the number of standard deviations a query point is away from the mean of a distribution in each dimension that is used to describe the data.

These methods have also been adapted to artificial neural networks (including deep belief networks) where the distance of the query point from the distribution is measured in the latent space corresponding to the final hidden layer \cite{janet2019quantitative}. A quantitative comparison of several methods including those described above are detailed in Refs.\ \cite{dnn_uncertainty, tran2020methods, hirschfeld2020uncertainty, ijms21155542}.

\section{Concluding Remarks}
\label{sec:conclusions}
In the development and application of ML models, much attention is paid to issues such as the choice of feature representation, data preprocessing, and model selection. While these are all important issues, this review highlights error analysis techniques and metrics as another vital part of ML workflows. 
The presented analyses and metrics allow us to validate ML models and assess their quality, reliability, and applicability. They also provide the foundation for model development, model comparison, model optimization, and the establishing of guidelines for the deployment of ML in the chemistry domain.  
Even sophisticated ML models that are trained on very large datasets can easily fail when used without careful consideration of their limitations, and such limitations need to be reported so that potential users are aware of them.
The discussed metrics can serve this purpose by illuminating different aspects of the performance of ML models and thus insuring that ML is in a position to advance chemical and materials domain knowledge.
The issue of metrics is crucial to further democratize the use of ML in the chemistry community, to promote best practices, to contextualize prediction results and methodological developments, and more broadly to instill the scientific outputs derived from ML work with trust, legitimacy, and transparency. 
(See also Outstanding Questions.)

\section*{Highlights}
\label{sec:highlights}
\begin{itemize}
	\item As machine learning (ML) is gaining an increasingly prominent role in chemical research, so is the need to assess the quality and applicability of ML models, compare different ML models, and develop best-practice guidelines for their design and utilization. 
Statistical loss function metrics and uncertainty quantification techniques are key issues in this context. 
	\item Different analyses highlight different facets of a model's performance, and a compilation of metrics -- as opposed to a single metric -- allows for a well-rounded understanding of what can be expected from a model. 
They also allow us to identify unexplored regions of chemical space and pursue their survey. 
	\item Metrics can thus make an important contribution to further democratize ML in chemistry, promote best practices, provide context to predictions and methodological developments, lend trust, legitimacy, and transparency to results from ML studies, and ultimately advance chemical domain knowledge. 
\end{itemize}

\section*{Outstanding Questions}
\label{sec:questions}

How can we facilitate a greater awareness, appreciation, and education of statistical techniques as well as data science more broadly? 
It has become clear that there is a need to update traditional curricula in the chemistry domain to account for its rapidly changing research landscape. 
It has also become clear that these analyses need to be incorporated into ML software tools as prominent features, \eg\ as part of automated recommender systems.

How can we expand our notion of benchmarking and error analysis to put a stronger emphasis on cost-benefit analysis? 
Given the increasing complexity of ML models that greatly increase their computational demand, it is worth asking if these efforts are actually worthwhile, in particular if they only lead to marginal improvements in the predictive performance.  

How can we advance the development and utilization of local rather than global error metrics?
Errors are generally not homogeneously distributed across all predictions but will differ in different prediction domains. As not all prediction domains are of equal interest, it is important to further advance methods that can gauge the quality of predictions where we are primarily interested in them. In chemical applications, this is often in extreme (potentially extrapolative) value regions with sparse training data and significantly larger than average errors.

How can we better harness our knowledge of the mathematical structure of different ML models to contextualize the specific error behavior of these models? Can we correlate the nature of latent variables, parametrization, model robustness, \textit{etc.}, with the predictive performance of the corresponding ML model?

\section*{Metric Equations}

\begin{center}
    \underline{Metrics for Regression Tasks}
\end{center}

\begin{equation} \tag{SI1}
\label{ei}
    e_{i} = y_{i, true} - y_{i, pred}
\end{equation}

\begin{equation} \tag{SI2}
\label{ri}
    r_{i} = \frac{y_{i, true} - y_{i, pred}}{y_{i, true}}
\end{equation}

\begin{equation} \tag{SI3}
\label{ybar}
    \Bar{y} = \frac{1}{n} \sum_{i=1}^{n} y_{i}
\end{equation}

\begin{equation} \tag{SI4}
\label{mae}
    MAE = \frac{1}{n} \sum_{i=1}^{n} \mathopen| e_{i} \mathclose|
\end{equation}

\begin{equation} \tag{SI5}
\label{rmse}
    RMSE = \sqrt{ \frac{1}{n} \sum_{i=1}^{n}  {\mathopen| e_{i} \mathclose|}^2 }
\end{equation}


\begin{equation} \tag{SI6}
\label{mape}
    MAPE = \frac{1}{n} \sum_{i=1}^{n} \left\lvert r_{i} \right\rvert \; \cdot 100\%
\end{equation}

\begin{equation} \tag{SI7}
\label{rmspe}
    RMSPE = \sqrt{ \frac{1}{n} \sum_{i=1}^{n}  {\left\lvert r_{i}\right\rvert}^2 } \; \cdot 100\%
\end{equation}

\begin{equation} \tag{SI8}
\label{me}
    ME = \frac{1}{n} \sum_{i=1}^{n} e_{i}
\end{equation}

\begin{equation} \tag{SI9}
\label{mpe}
    MPE = \frac{1}{n} \sum_{i=1}^{n} r_{i} \; \cdot 100\%
\end{equation}

\begin{equation} \tag{SI10}
\label{maxae}
    MaxAE = max\{ \, e_{i} \, \}, \quad  i = 1,...,n
\end{equation}

\begin{equation} \tag{SI11}
\label{maxape}
    MaxAPE = max\{ \, \mathopen| r_{i} \mathclose| \; \cdot 100\% \, \}, \quad  i = 1,...,n
\end{equation}

\begin{equation} \tag{SI12}
\label{deltamaxe}
    \Delta MaxE = max\{ \, e_{i} \, \} - min\{ \, e_{i} \, \} , \quad  i = 1,...,n
\end{equation}

\begin{equation} \tag{SI13}
\label{sigma}
    \sigma = \sqrt{\frac{1}{n} \sum_{i=1}^{n}  ({y_{i} - \Bar{y}})^2}
\end{equation}

\begin{equation} \tag{SI14}
\label{r2}
    R^2 = 1 - \sum_{i=1}^{n} \frac{ {\mathopen| e_{i} \mathclose|}^2}{{\mathopen| y_{i, true} - \Bar{y} \mathclose|}^2} 
\end{equation}

\begin{equation} \tag{SI15}
\label{adjr2}
    R^2_{adj} = 1 - \frac{(n - 1)}{(n - m - 1)} \sum_{i=1}^{n} \frac{ {\mathopen| e_{i} \mathclose|}^2}{{\mathopen| y_{i, true} - \Bar{y} \mathclose|}^2}
\end{equation}

\begin{center}
    \vskip 0.5cm
    \underline{Metrics for Classification Tasks}
\end{center}

\begin{equation} \tag{SI16}
\label{acc}
    Acc = \frac{TP + TN}{TP + FP + FN + TN}
\end{equation}

\begin{equation} \tag{SI17}
\label{prec}
    Prec = \frac{TP}{TP + FP}
\end{equation}

\begin{equation} \tag{SI18}
\label{rec}
    Rec = TPR = \frac{TP}{TP + FN}
\end{equation}

\begin{equation} \tag{SI19}
\label{f1}
    F1 = 2 \cdot \frac{Prec \cdot Rec}{Prec + Rec}
\end{equation}

\begin{equation} \tag{SI20}
\label{wf1}
    F1_\beta = (1+\beta) \cdot \frac{Prec \cdot Rec}{\beta^2 \cdot Prec + Rec}
\end{equation}

\begin{multline} \tag{SI21}
\label{logloss}
\quad \quad \quad \quad \quad \mathcal{L} = - \log\ P(y_{i, true}\mathopen|y_{i, pred}) \\ = -(y_{i, true}\ \log(y_{i, pred})\ + \\ (1 - y_{i, true})\log(1 - y_{i, pred})    
\end{multline}

\begin{equation} \tag{SI22}
\label{fpr}
    FPR = \frac{FP}{FP + TN}
\end{equation}

\begin{equation} \tag{SI23}
\label{tanimoto}
    T = \frac{w}{u + v - w}
\end{equation}

\begin{align*}
with: \\
Acc &: Accuracy, \\ 
Prec &: Precision, \\ 
Rec &: Recall, \\ 
n &: total\ no.\ of\ data\ points, \\
m &: total\ no.\ of\ features, \\
u &: total\ no.\ of\ features\ in\ 1st\ molecule, \\ 
v &: total\ no.\ of\ features\ in\ 2nd\ molecule, \\ 
w &: no.\ of\ common\ features\ between\ the\ 2\ molecules 
\end{align*}

\section*{Glossary}

\noindent 
\textit{Binary cross-entropy}: In a binary classification problem, each sample belongs to either one class or the other, \ie\ it has a known probability of 1.0 for one class and 0.0 for the other. 
A classifier model can estimate the probability of a sample belonging to each class. 
The binary cross-entropy is used as a metric to assess the difference between the two probability distributions 
and thus the uncertainty of a classifier's prediction. (Also see \textit{cross-entropy}, \textit{categorical cross-entropy}, and \textit{log loss}.)

\vskip 0.5cm

\noindent 
\textit{Categorical cross-entropy}: For multi-class classification problems, \ie\ for problems involving more than two categories (classes) of data, the \textit{cross-entropy} measures the difference between the probability distribution of a sample belonging to one class, and the probability distribution of that sample not belonging to that class, \ie\ belonging to any of the other classes. This metric is known as categorical cross-entropy. (Also see \textit{binary cross-entropy}.)

\vskip 0.5cm

\noindent 
\textit{Cross-entropy}: This is a measure of the difference between two probability distributions for a given set of samples. (Also see \textit{binary cross-entropy}, \textit{categorical cross-entropy}, and \textit{log loss}.)

\vskip 0.5cm

\noindent 
\textit{Evolutionary/genetic algorithm}: This is a heuristic-based approach inspired by natural selection in biological processes (\ie\ survival of the fittest). It is typically employed to tackle (combinatorial) optimization problems, in which gradients (needed for gradient descent methods) are ill defined (\eg\ in problems involving discrete or categorical variables) or otherwise inaccessible. Each possible solution behaves as an individual in a population of solutions and a \textit{fitness function} -- itself a \textit{loss function metric} -- is used to determine its quality. Evolutionary optimization of the population takes place \via\ reproduction, mutation, crossover, and selection iterations. 

\vskip 0.5cm

\noindent 
\textit{Fitness/objective function}: This is a \textit{loss function metric} that assesses the quality of a solution with respect to an objective of an optimization. Its output can be maximized or minimized, \eg\ as part of an \textit{evolutionary algorithm}.

\vskip 0.5cm

\noindent 
\textit{Harmonic mean}: This is one of multiple types of mean value metrics. Given a set of sample values, the harmonic mean is the inverse of the arithmetic mean of the inverse of the sample values. 

\vskip 0.5cm

\noindent 
\textit{Hyperparameters}: In ML, hyperparameters are the parameters that define the structure of a model and control the learning process, as opposed to other parameters that are derived (`learned') from the data in the course of training the model.

\vskip 0.5cm

\noindent 
\textit{Log loss}: This is the negative logarithm of the likelihood of a set of observations given a model's parameters. While log loss and \textit{cross-entropy} are not the same by definition, they calculate the same quantity when used as \textit{fitness functions}. In practice, the two terms are thus often used interchangeably.

\vskip 0.5cm

\noindent 
\textit{Loss function metrics}: These are statistical error metrics used to assess the performance of ML models and the quality of their predictions.

\vskip 0.5cm

\noindent 
\textit{Principal component analysis}: This is a technique to transform the feature basis, in which a set of data is described, into a basis that is adapted to the nature of the given data. The principal components are the eigenvectors of the covariance matrix of the data set.

\vskip 0.5cm

\noindent 
\textit{Tanimoto index}: This metric is used to assess the similarity between the finite feature (\eg\ descriptor, fingerprint) vectors of two samples. The similarity ranges from 0 to 1, with 0 indicating no point of intersection between the two vectors, and 1 revealing completely identical vectors.

\vskip 0.5cm

\section*{Competing Financial Interests}
The authors declare to have no competing financial interests.

\begin{acknowledgments} 
This work was supported by the NSF CAREER program under grant No.\ OAC-1751161 and the NSF Big Data Spokes program under grant No.\ IIS-1761990.  
\end{acknowledgments}

\bibliographystyle{custom}
\bibliography{54_benchmarking_metrics}
\end{document}